\documentclass[%
groupedaddress,
 preprint,
 aps,
]{revtex4-1}

\usepackage{graphicx}
\usepackage{dcolumn}
\usepackage{bm}
\usepackage{lineno}

\newcommand{\hide}[1]{} 

\usepackage{amsmath}            
\usepackage{amssymb}            
\usepackage{latexsym}
\usepackage{subfigure}
\usepackage{rotating}
\usepackage{graphicx}
\usepackage{float}

\newcommand{\eg}{e.g., }

\newcommand{\ie}{i.e., }

\newcommand{\vs}{vs.\ }

\newcommand{\fig}[1]{Fig.\ \ref{fig:#1}}
\newcommand{\Fig}[1]{Figure \ref{fig:#1}}

\newcommand{\tbl}[1]{Table \ref{t:#1}}

\begin{document}
\graphicspath{{Figs/}}

\title{Granular Convection in Microgravity}

\author{N.~Murdoch$^{1,2}$}
	\email{murdoch@oca.eu}
\author{B.~Rozitis$^2$}
\author{K.~Nordstrom$^3$}
\author{S.F.~Green$^2$}
\author{P.~Michel$^1$}
\author{T-L.~de Lophem$^2$} 
\author{W.~Losert$^3$}
   \email{wlosert@umd.edu}
\affiliation{$^1$Laboratoire Lagrange, UMR 7293, Universit\'{e} de Nice Sophia-Antipolis,  CNRS, Observatoire de la C\^{o}te d'Azur, 06300 Nice, France}
\affiliation{$^2$Planetary and Space Sciences, Department of Physical Sciences, The Open University, Milton Keynes, UK}
\affiliation{$^3$Institute for Physical Science and Technology, and Department of Physics, University of Maryland, USA }

\date{Published in Physical Review Letters on 3rd January 2013}

\begin{abstract}
We investigate the role of gravity on convection in a dense granular shear flow. Using a microgravity-modified Taylor-Couette shear cell under the conditions of parabolic flight microgravity, we demonstrate experimentally that secondary, convective-like flows in a sheared granular material are close to zero in microgravity, and enhanced under high gravity conditions, though the primary flow fields are unaffected by gravity. We suggest that gravity tunes the frictional particle-particle and particle-wall interactions, which have been proposed to drive the secondary flow. In addition, the degree of plastic deformation increases with increasing gravitational forces, supporting the notion that friction is the ultimate cause.
\end{abstract}

\pacs{45.70.-n  47.57.Gc  64.60.aq  64.60.ah}

\maketitle


Characterising and predicting flow of granular materials in response to shear stress is an important geophysical and industrial  challenge.  Granular flow has been studied in depth \citep[\eg][]{losert00, cheng06, toiya04, fan10}, often using Taylor-Couette shear cells, where shear stress is applied between two concentric cylinders.  This leads to strain fields between the cylinders, and generally localised shear bands.  In addition, several shear cell experiments found convective-like motion near the shear zone \citep[\eg][]{khosropour97}.  This secondary flow is considered key for important practical processes such as segregation \citep{khosropour97}. The wealth of experimental evidence demonstrating that convective flows can occur in a granular material \citep[\eg][]{rodriguez06, eshuis10, liffman97, tennakoon99} has led to significant theoretical effort, with a number of proposed mechanisms for granular convection \citep[\eg][]{taguchi92,rajchenbach91,shinbrot97,rodriguez06,cordero03}.   Gravity is considered as a potential driving force in some of the models \citep{ramirez00,forterre01, rajchenbach91,gray99}. 

Using a parabolic flight environment to study the dynamics of granular material subject to shear forces in a Taylor-Couette shear cell, we investigate the role of gravity in driving secondary flows within a dense confined granular flow.

\emph{Experimental set-up and procedures. }  Our experiments use a Taylor-Couette geometry.  There are two concentric cylinders. The outer cylinder is fixed and its inside surface is rough with a layer of particles, and the outer surface of the inner cylinder is also rough and rotated to generate shear strain.  The floor between the two cylinders is smooth and fixed in place. The gap between the two cylinders is filled, to a height of 100 mm, with spherical soda lime glass beads (grain diameter, $d$ $=$ 3 mm; density, $\rho$ $=$  2.55 g cm$^{-3}$) upon which the rotating inner cylinder applies shear stresses. A movable and transparent disk is used to confine the granular material during the microgravity phase of a parabola with an average force of 6.6 N (the force can vary from 0 to 13.2 N depending on the packing fraction of the granular material).

During each parabola of a parabolic flight there are three distinct phases: a 20 second $\sim$1.8 $g$ (where $g$ is the Earth's gravitational acceleration) injection phase as the plane accelerates upwards, a 22 second microgravity phase ($\sim \pm 10^{-2} g$) as the plane flies on a parabolic trajectory (during this period the pilot carefully adjusts the thrust of the aircraft to compensate for the air drag so that there is no lift) and, lastly, a 20 second $\sim$1.8 $g$ recovery phase as the plane pulls out of the parabola. 

\begin{figure}[h]
\centering
\subfigure[]{
\centering
\includegraphics[width=0.45\columnwidth]{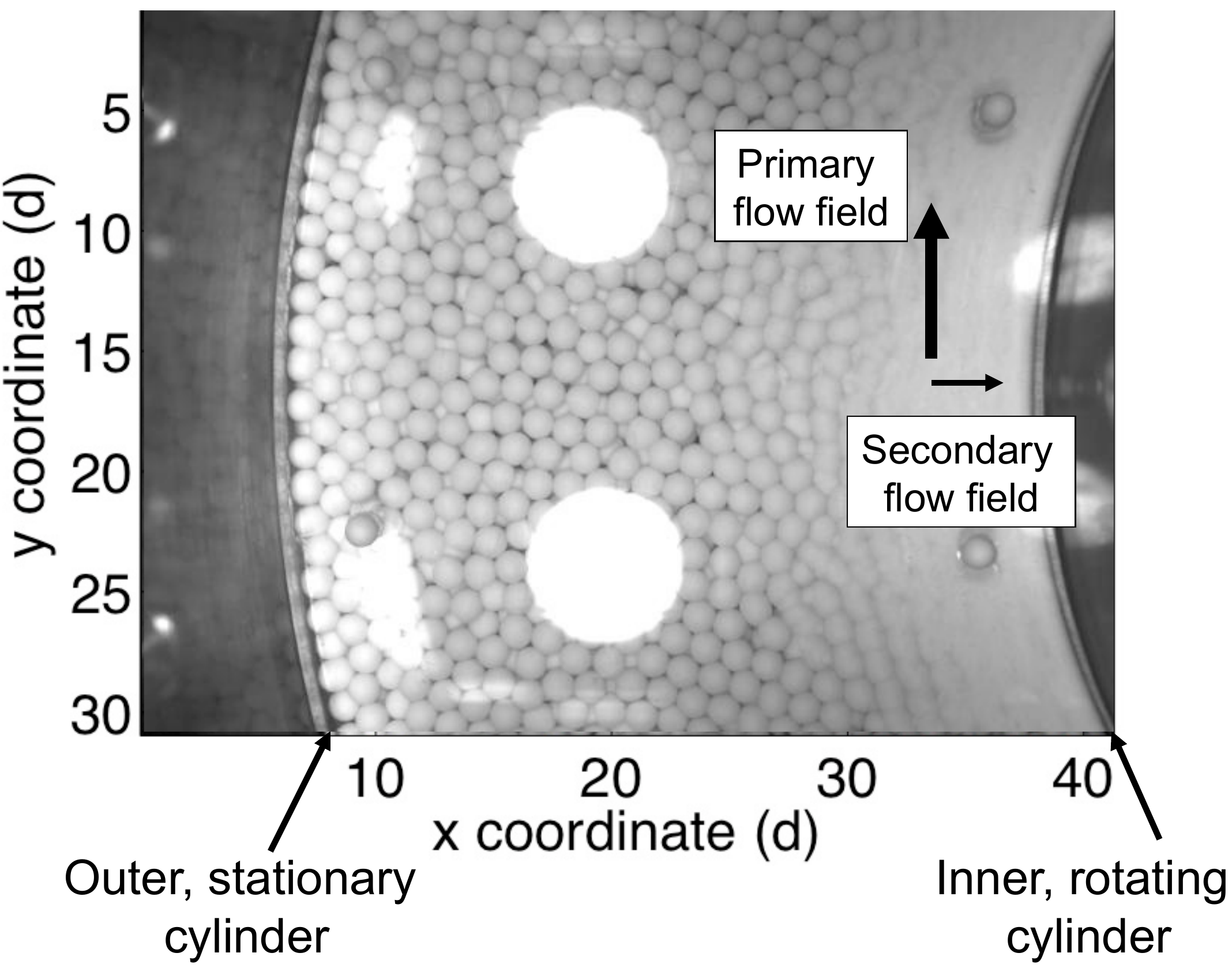}
 \label{fig:Trajectories}
}
\subfigure[]{
\centering
\includegraphics[width=0.48\columnwidth]{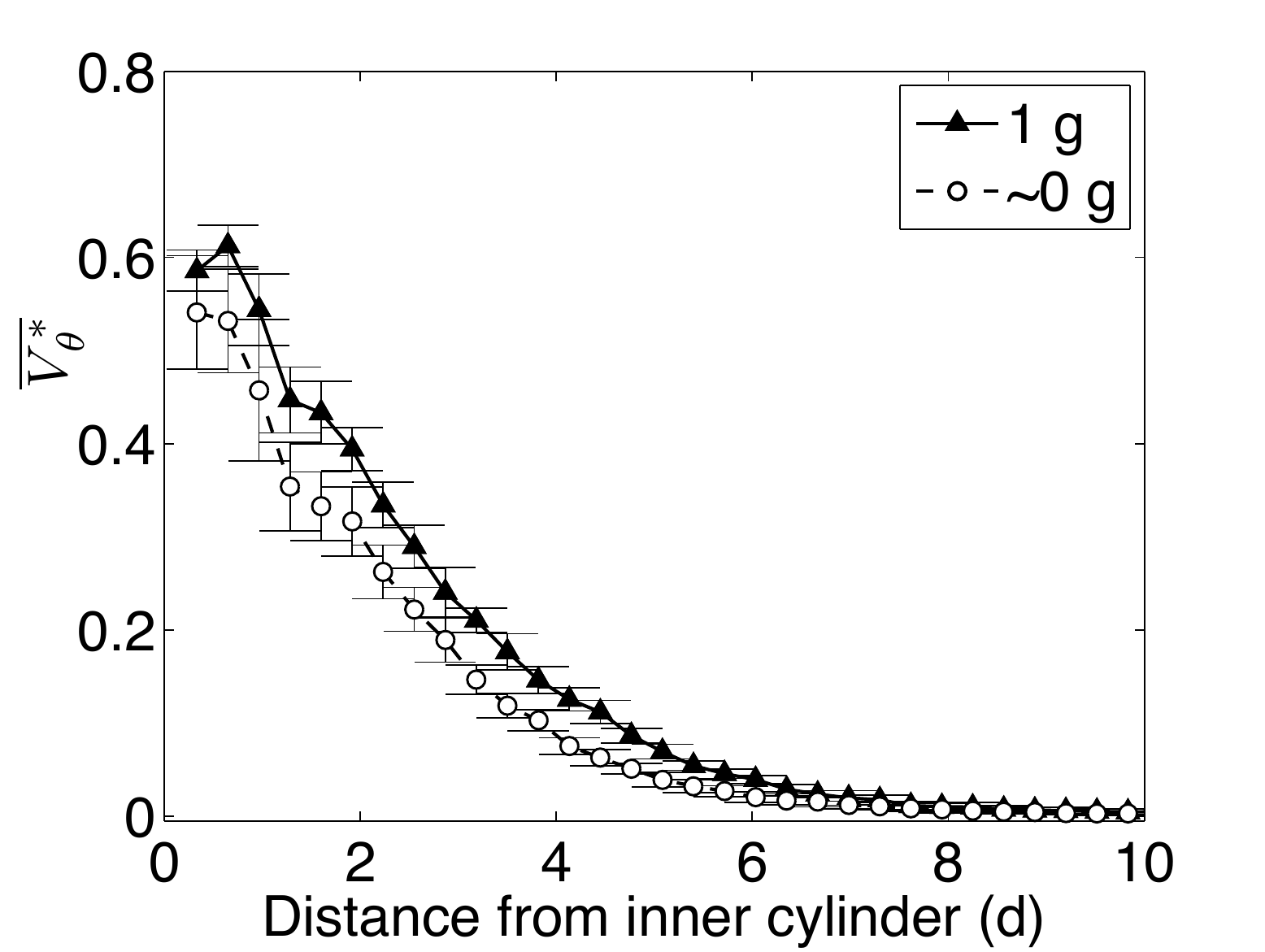}
\label{fig:Cam2_3mm_MeanAngVelProfs}
}
\caption{(a) Superposition of experimental images showing the particle motion during $\sim$60 seconds of a ground-based experiment. The bright areas in the image are reflections of the lamps on the confining pressure plate and cylinder walls. The four beads that are glued to the top surface of the confining plate (to determine the pixel scale) can also be seen. The primary and secondary flow fields are shown. Close to the inner, rotating cylinder the magnitude of the primary flow field is $\sim$0.6 $\omega$. (b) Comparison of angular velocity profiles of the particles in 1 $g$ and low-gravity.  $\overline{V^*_{\theta}}$ ($\overline{V^*_{\theta}} =  \frac{\overline{V_{\theta}}}{\omega}$ where $\overline{V_{\theta}}$ is the mean angular velocity of several experiments of the same type) plotted as a function of distance from the inner cylinder for the top surface of ground-based and microgravity experiments. The error bars represent the standard deviation of $\overline{V^*_{\theta}}$ for each group of experiments. The velocity profiles shown only extend up to 10 $d$.}
\end{figure}

The motor that drives the inner cylinder was started shortly after the microgravity phase begins for each parabola and ran until the 1 $g$ rest phase started. High-speed cameras imaged the top and bottom layers of glass beads in the shear cell at $\sim$60 frames sec$^{-1}$ so that the particles did not move more than 1/10 $d$ between consecutive frames. \Fig{Trajectories} is a stacked image of one experiment showing the particle motion. Experiments were performed with the inner cylinder rotating at 0.025, 0.05 and 0.1 rad sec$^{-1}$.  In between the parabolas the shear cell is shaken by hand to attempt to reproduce the same initial bulk packing fraction while minimising possible memory effects from prior shear.  Further details of our experimental design can be found in the Supplementary Material and \cite{murdoch12_IEF}.

After the flights particle tracking was performed using an adaptation of a subpixel-accuracy particle detection and tracking algorithm \citep{crocker96}, which locates particles with an accuracy of approximately $1/10$ pixel. The raw particle position data was smoothed over time using a local regression weighted linear least squares fit. From this, the average particle velocities were computed. 
 
We have found that, between the gravitational regimes of microgravity and 1 $g$, there is no difference in the width of the shear band nor is there a large difference in the magnitude of the angular (tangential) velocities within the shear band \citep[see \fig{Cam2_3mm_MeanAngVelProfs} and ][]{murdoch12_IEF}.  The primary flow field exhibits shear banding, consistent with prior work in this geometry.  Shear banding has been shown to be insensitive to loading at the particle contacts \citep{losert00} and substantial changes the geometry of particles \citep{mueth00}.  Our observed insensitivity of the primary flow to changes in gravity may also be due to the fact that both the primary flow direction as well as the shear gradient direction are perpendicular to gravity.

\begin{figure}[h]
  \centering
\includegraphics[width=0.75\columnwidth]{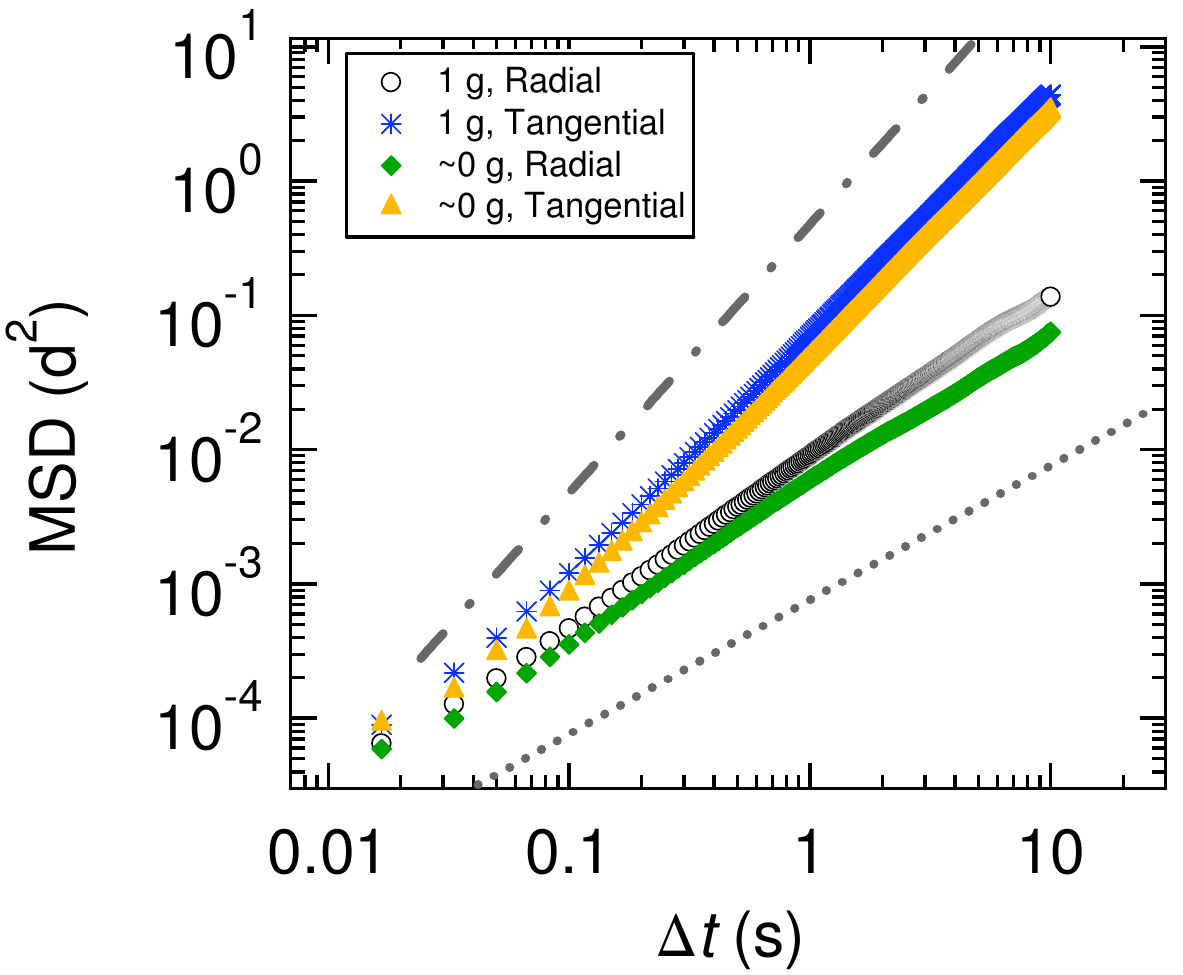}
 \caption{Mean-square displacement (MSD) of particles in microgravity and on the ground (for particles with $(r - a) <6.5 d$).  The four curves represent the tangential and radial MSD for the ground and for microgravity. The dotted line shows a slope of 1, the dotted-dashed line shows a slope of 2. In these experiments $\omega$ = 0.025 rad sec$^{-1}$.}
  \label{fig:MSDs}
\end{figure}

We find that there is also very little difference in the particle mean-square displacement (MSD) in the tangential direction between the ground and microgravity experiments; in both cases tangential MSDs indicate close to ballistic motion (a power law fit of MSD \vs time yields an exponent of 1.8; see \fig{MSDs}). The MSDs of the ground and microgravity experiments are also very similar in the radial direction; both experiments show displacements slightly greater than expected for purely diffusive motion (power law exponent of 1.1; see \fig{MSDs}). The power law exponent in the tangential direction is consistent with previous experimental observations in a 2-d system \citep[\eg][]{utter04}. However, the power law in the radial direction is slightly less than 1 in the 2-d system, indicating subdiffusive motion, while it is greater than 1 in our measurements in 3-d.  This suggests additional drift in the radial direction in our 3-d system consistent with convective flow.

 \vspace{0.1cm}

\begin{figure}[h]
\centering
 \includegraphics[width=0.75\columnwidth]{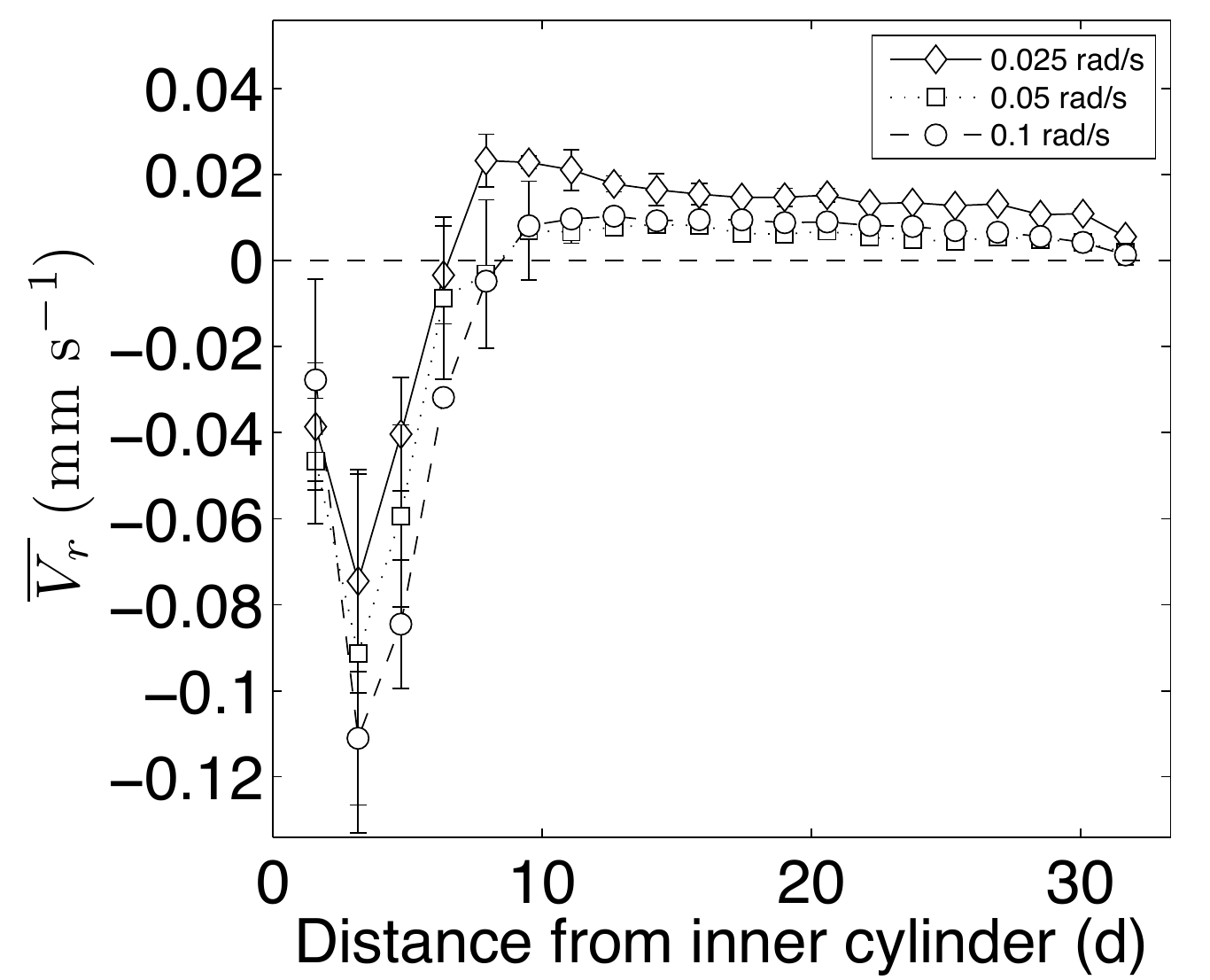}
 \put (-210,45){ \includegraphics[width=0.4\columnwidth]{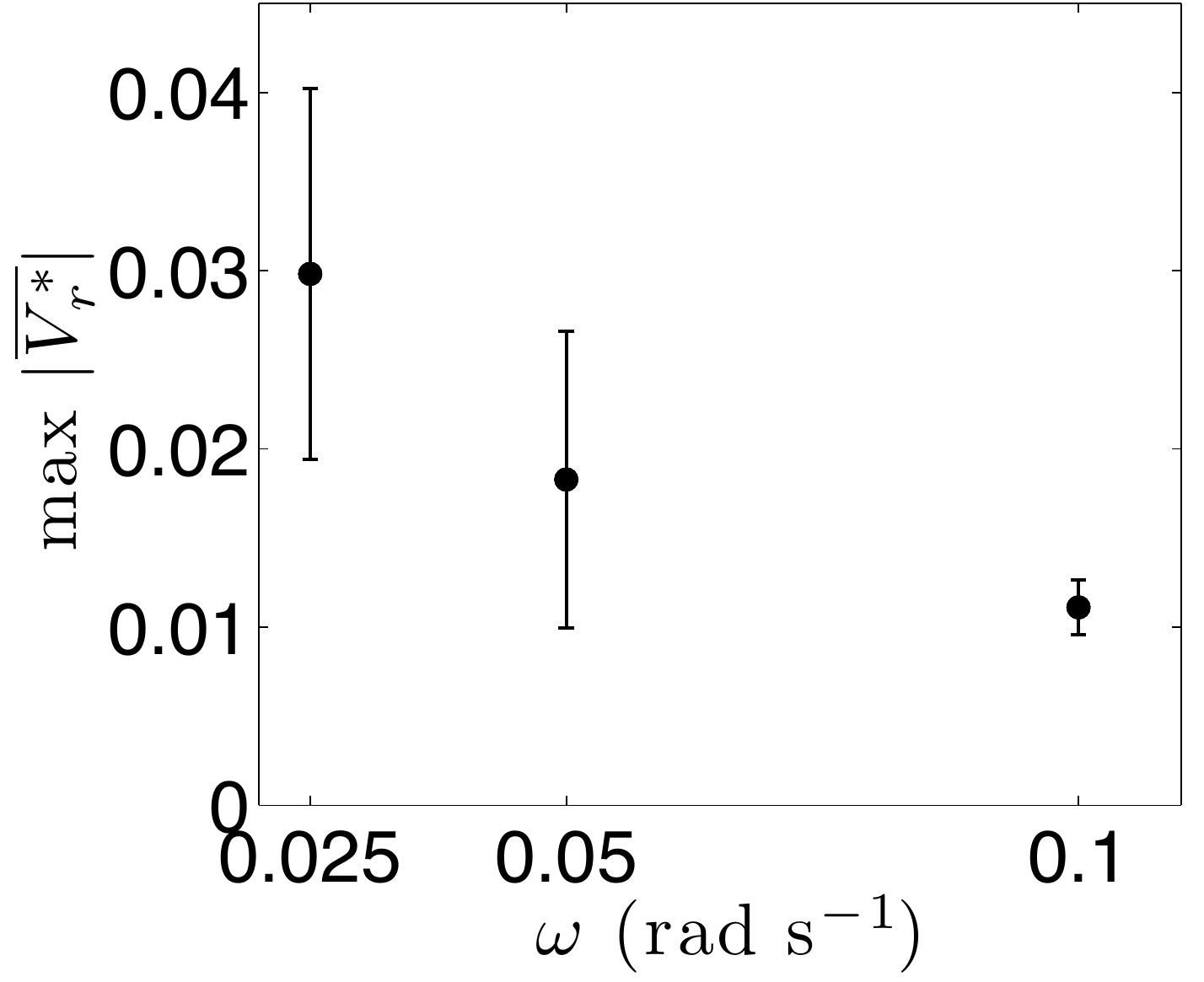}}
 \caption{Mean radial velocity ($\overline{V_r}$) is shown as a function of distance from the inner cylinder on the top surface of experiments for ground-based experiments with different inner cylinder angular velocities. The error bars represent the standard deviation of $\overline{V_r}$ for each group of experiments. Inset: Magnitude of the maximum normalised radial velocity ($\overline{V^*_{r}}= \frac{\overline{V_{r}}}{(a \omega)}$) as a function of $\omega$.}
  \label{fig:Cam2_3mm_1gRadProfs}  
\end{figure}

\emph{Convective particle motion. } Another indication of convective motion comes from the radial velocity profiles. Specifically, since the packing density of the granular material is approximately constant everywhere, any radial inward motion on the top surface must be compensated with radial outward motion below the surface, indicating a likely convective flow.  \Fig{Cam2_3mm_1gRadProfs} shows the radial velocity profiles as a function of distance from the inner cylinder, for the top surface of a set of experiments at normal gravity with different inner cylinder angular velocities. Although observations vary from experiment to experiment, there is a reproducible trend in the shape of the radial velocity profiles. All of the ground-based experiments exhibit a region of negative radial velocity, which approximately coincides with the shear band (see \fig{Cam2_3mm_MeanAngVelProfs}). Despite the small scale of the radial motion ($0.2\%$ of the tangential motion at the inner cylinder, more at the outer edge of the shear band), there is clearly a preferred radial direction of particle motion in this region of the top surface for all ground-based experiments. 

We also tracked particle motion at the bottom surface of the shear cell.  The primary flow field is comparable to the top surface. At normal gravity, the scale of the radial motion on the bottom surface is much smaller than that on the top surface and appears random.  This indicates that the convective flow at normal gravity does not extend all the way to the bottom of the shear cell. 

The radial flow speed on the top surface increases with increasing primary flow speed (\fig{Cam2_3mm_1gRadProfs}), indicating that the convective flow speed depends on the primary flow speed.  However, the magnitude of the radial velocity normalised by shear rate decreases with increasing shear rate (see inset of \fig{Cam2_3mm_1gRadProfs}). This observation indicates that convective flow is not only driven by rearrangements that are needed for shear, but that the rearrangements responsible for convective flow have an independent timescale.  This observation is consistent with a gravity driven convective flow field. 

\begin{figure}[h]
 \centering
\includegraphics[width=0.75\columnwidth]{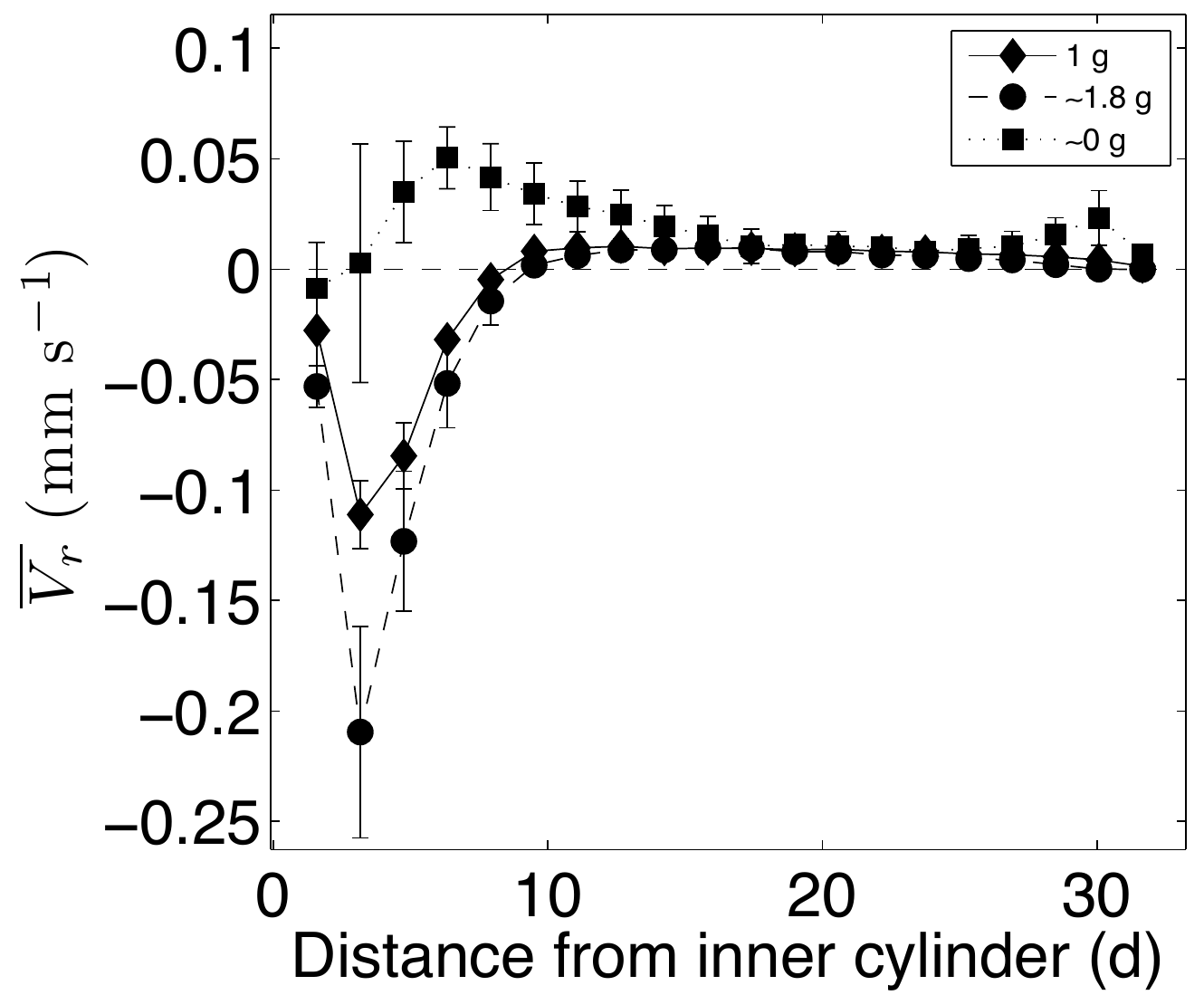}
 \put (-210,45){\includegraphics[width=0.42\columnwidth]{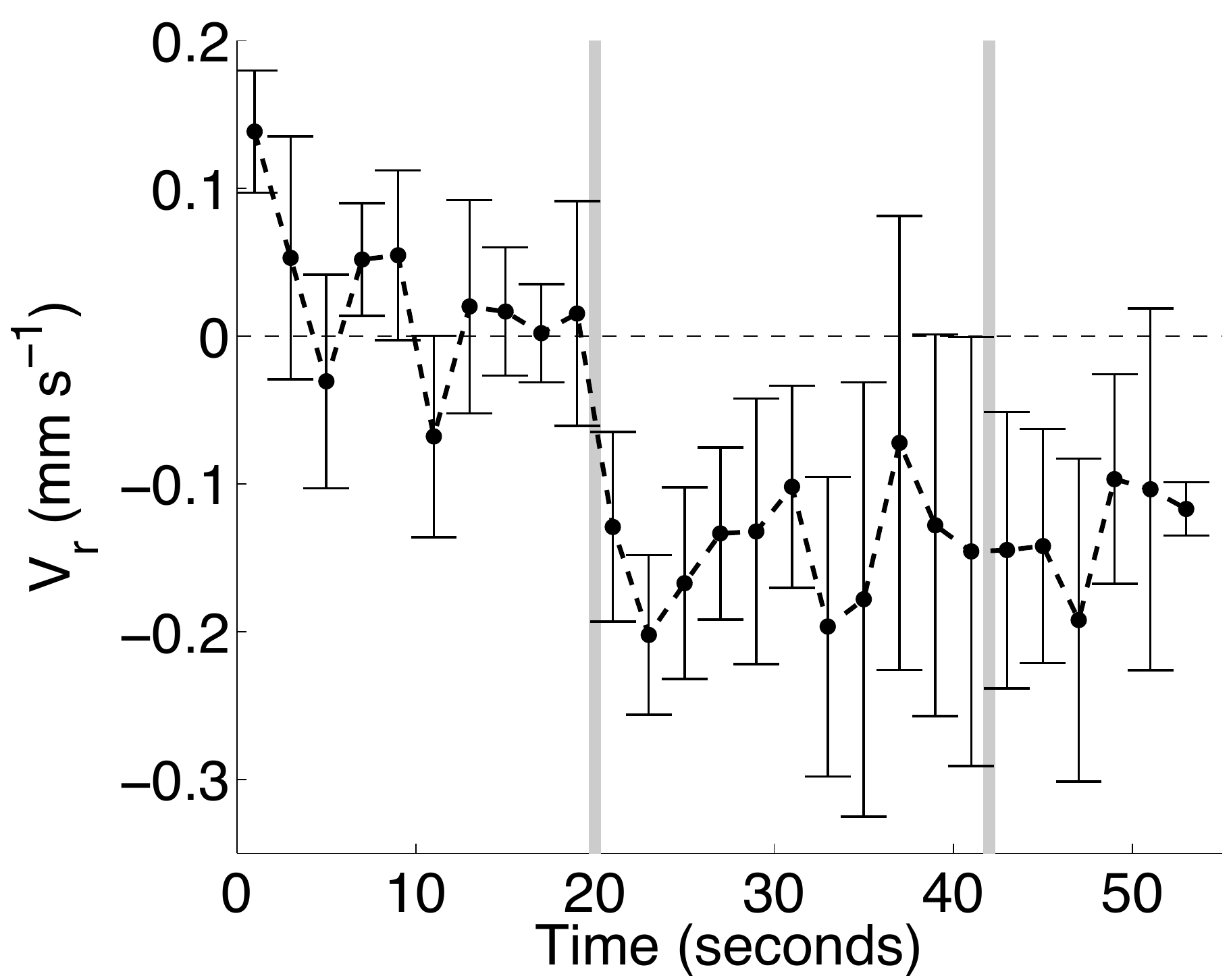}}
\put (-155, 90) {$\sim$0 $g$}
\put (-100, 180) {$\sim$1.8 $g$}
\put (-40, 180) {1 $g$}
 \caption{Mean radial velocity ($\overline{V_r}$) as a function of distance from the inner cylinder on the top surface for microgravity, ground-based and $\sim$1.8 $g$ experiments. The error bars represent the standard deviation of $\overline{V_r}$ for a group of experiments with $\omega$ = 0.1 rad sec$^{-1}$ (5 experiments for microgravity and $\sim$1.8 $g$, 2 for ground-based experiments). Inset: ${V_r}$ \vs time for particles with $(r - a) <6.5 d$.  Gravitational forces are not present from 0 - 20 secs then, from 20  - 42 secs, the gravitational acceleration is $\sim1.8$ $g$ and, finally, after 42 secs the gravitational acceleration returns to $\sim1$ $g$. Each point represents the mean $V_r$ of the particles in the radial bin during a period of 2 seconds and $\omega$ = 0.1 rad sec$^{-1}$.  The error bars represent the standard deviation of $V_r$ in each 2 second period. }
\label{fig:VaryingGrav}
\end{figure}

\Fig{VaryingGrav} shows how the mean radial velocity profiles change as a function of gravitational acceleration. The magnitude of the negative radial velocities in $\sim$1.8 $g$ is larger than in the ground-based experiments. Conversely, in microgravity there is no inward particle motion.  Thus there is a correlation between gravity and secondary flows. 

We investigate causality by looking at time traces of the radial velocity; the sample experiences three distinct gravity values during a flight. The change in the particle dynamics occurs rapidly when the gravitational environment changes. The inset of \fig{VaryingGrav} shows the mean radial velocity as a function of time during a transition from microgravity to gravity. Note that the radial inward flow starts very quickly, within less than 2 seconds (the slow radial flow speeds are not detectable on shorter timescales). Clearly, we turn on the convective flow by flipping the gravity switch to ``on''. This suggests that gravity is indeed causing the flow.

This begs the question of how gravity causes secondary flows. It has been shown experimentally that friction plays a deciding role in whether secondary flows occur in another flow geometry \citep{clement92}. We conjecture that in our system, gravity acts as an amplifier for frictional effects. 

A granular bed under gravity is supported by its constituent particles; top grains are supported by bottom grains, which are ultimately supported by the bottom and sidewalls. Force chains of contacting particles bear the brunt of load, transmitting the force to the bottom and sidewalls. Particles in force chains have substantial normal, and thereby frictional, forces between them \citep{majmudar05}.  Gravity thus creates a vertical gradient of interparticle forces, resulting in particle rearrangements being more likely near the surface.

In a horizontal slice of the Couette cell, particles are more likely to rearrange near the inner cylinder due to the shear forces breaking their contacts. Under no gravity, the forces at the top and bottom of the slice are equal, so the motion is only in the plane. Under gravity, this shear force does not change, and so the primary flow in the plane is unaffected. However, gravity introduces an asymmetry: the contact forces at the top and bottom of the slice are now different, and produce a secondary flow pattern.  As the individual particle motions are biased by gravity, the likely average flow pattern is set. Particles at the top and near the cylinder are likely to go down, since that is the highest rearrangement zone. Rearrangements are suppressed as one goes down the pile as the forces between particles become stronger, resulting in smaller secondary flows near the bottom. Far away from the cylinder, a tiny upward flow balances the pile. For higher gravity, the interparticle and particle-boundary forces increase, and the vertical contact gradient is greater. This creates stronger secondary flows in higher gravity. In contrast, with microgravity, the strength of contacts does not vary with depth, and the contacts transmit, on average, a smaller force. 

To test these notions, we look at particle rearrangements in the system, focusing on irreversible, plastic events that are signatures of force chain breaking \citep{makse99}, and can make up the average convective flow. In accordance with the idea of frictional force chain breaking, we would expect higher degrees of plastic deformation near the inner cylinder, and for plastic deformation to increase with higher gravity.

\emph{Plastic deformation. } To measure the local plastic rearrangements of particles relative to their neighbours on the top surface of our shear cell, we use $D^2_{min}$ \cite{chen10}, :

\begin{equation}
D^2_{min,i}=\text{min}\lbrace\sum_j[\Delta\overline d_{ij}(t)-E_i\overline d_{ij}] ^2 \rbrace
\label{D2min}
\end{equation}

\begin{figure}
\centering
\subfigure[]{
\centering
\includegraphics[width=0.3\columnwidth]{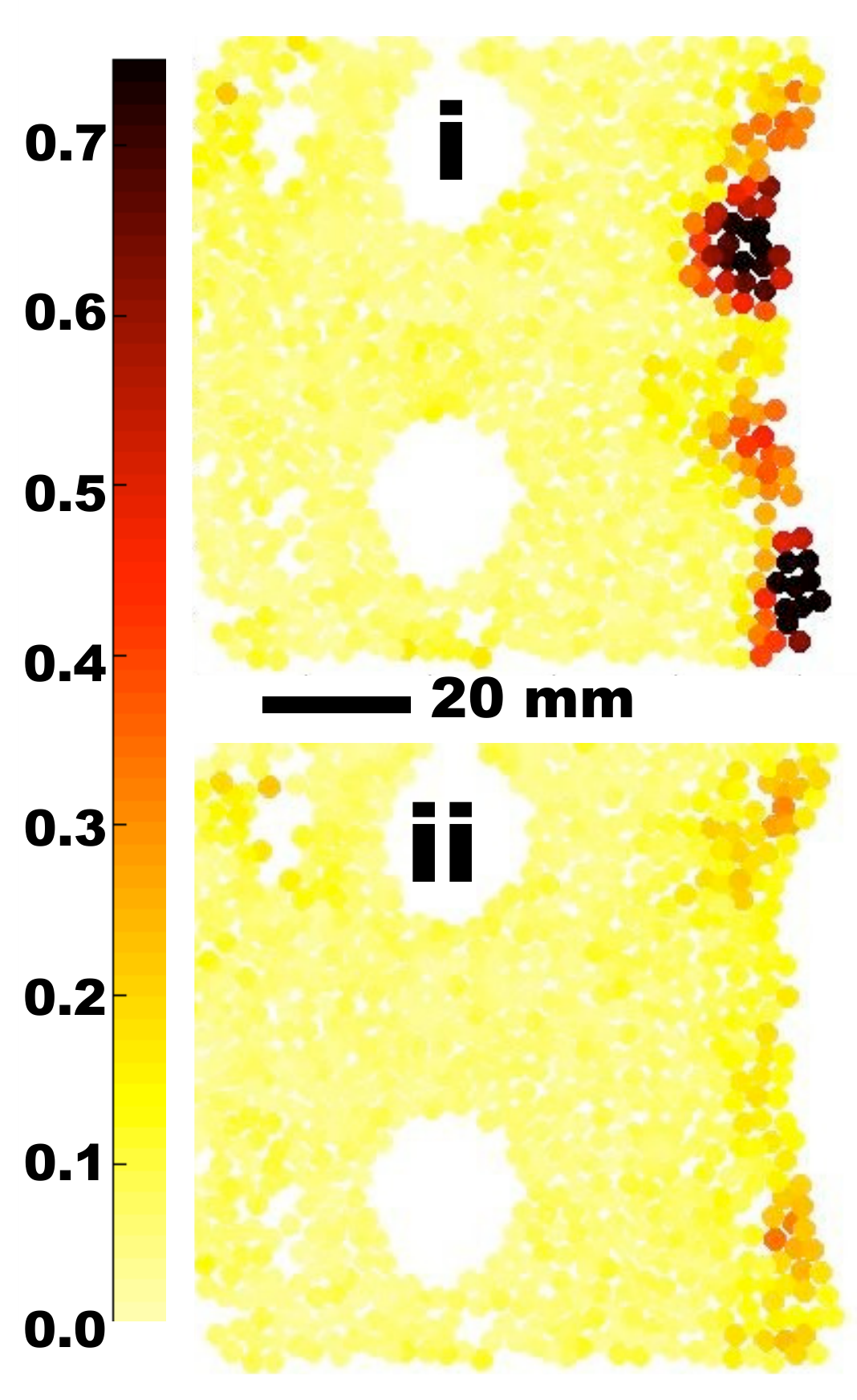}
  \label{fig:StillofD2min}
   }
 \subfigure[]{
 \centering
\includegraphics[width=0.62\columnwidth]{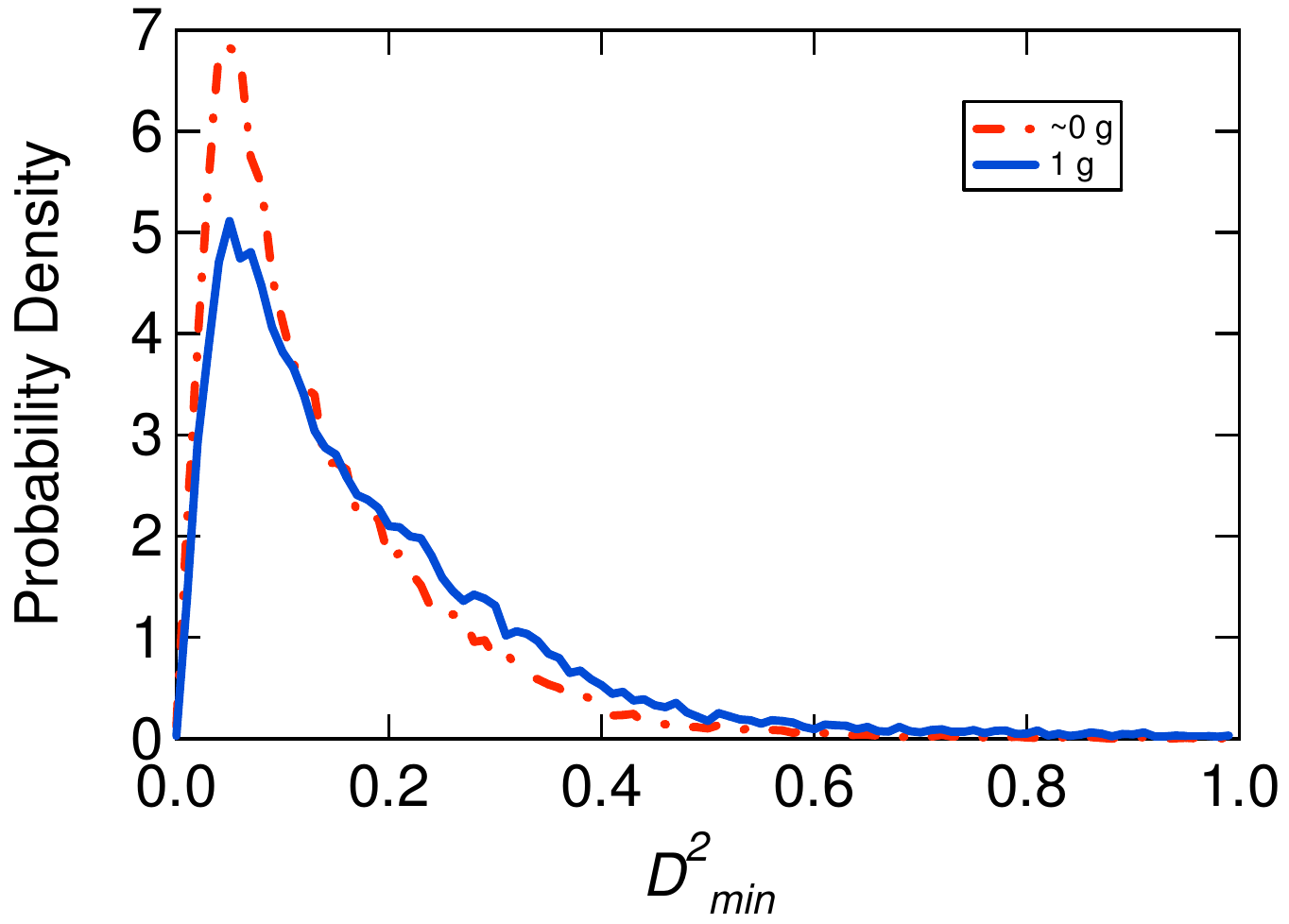}
  \label{fig:Histograms}
}
\label{fig:D2min}
 \caption{(a) Two snapshots of $D^2_{min}$ for a ground based experiment showing regions of large plastic deformation. Particles with a high and low $D^2_{min}$ are indicated in dark red and yellow, respectively (colour figure online). (b) Probability density of $D^2_{min}$ for particles with $(r - a) <6.5 d$ in the microgravity and ground-based experiments.}
\end{figure}

$D^2_{min,i}$ quantifies the nonaffine deformation of $j$ particles in the neighbourhood around a given particle $i$ after removing the averaged linear response to the strain, given by tensor $E_i$; a smaller $D^2_{min}$ indicates more affine motion. The vector $\overline d_{ij}$ is the relative position of $i$ and $j$, $\Delta\overline d_{ij}$ is the relative displacement after a delay time $\Delta t$, set to be 4 frames in this case. We normalise our values of $D^2_{min}$ by the tangential mean-square displacement at $\Delta t$, so that the magnitude of $D^2_{min}$ is not dependent on the macroscopic flow speed. Snapshots of $D^2_{min}$ values from a ground based experiment are shown in \fig{StillofD2min}. We find that the $D^2_{min}$ values are spatially heterogeneous, with high-valued ``hot spots'' near the inner cylinder both on the ground and in microgravity. These hot spots evolve for several frames, then die out after some plastic rearragement event. The average $D^2_{min}$ value \ie the average strength of the plastic deformation, is found to be larger on the ground (see \tbl{D2minValues} and \fig{Histograms}) compared to microgravity, and slightly larger still for the $\sim$1.8 $g$ case. This indicates that gravity does enhance this deformation. 

\vspace{0.2cm}

\emph{Conclusions. } 
We have shown that gravity plays an important role in the dynamics of a sheared dense granular flow. Radial flows (likely due to convection) are affected by gravity; they become larger in magnitude in the presence of increased gravitational acceleration, and disappear altogether in microgravity.  We suggest that gravity tunes the frictional particle-particle and particle-wall interactions, which have been proposed to drive the secondary flow.  Without a gradient in friction and with low friction, the secondary flow is halted.  To address the relative importance of the gradient in friction, future work should tune the frictional properties of the system, in addition to the confining pressure and gravity. We have shown that different frictional/normal gradients create different plastic deformations in the system. We believe plastic deformation, which may be different for the same primary flow field, could hold the key to understanding the driving forces of convection. Overall, while the primary flow field is deceivingly similar in normal conditions and microgravity, the absence/presence of gravity causes dramatic changes in secondary flow characteristics that are crucial in industrial applications, such as segregation by size, shape, and density, as well as astrophysical questions, such as understanding the behaviour of regolith on planetary surfaces.

We acknowledge support from the OU, TAS, STFC,
RAS, French PNP and ESA Fly your Thesis. WL and
KN acknowledge support from NSF Grant No. DMR-
0907146 and DTRA Grant No. 1-10-1-0021.

\begin{table}[h]
\caption{\label{t:D2minValues} $D^2_{min}$ values in the inner radial bin ($<$ 6.5 $d$ from rotating inner cylinder) for the three different gravity regimes.}
\begin{ruledtabular}
\begin{tabular}{cccc}
Gravitational  & Average $D^2_{min}$ & $\sigma/\sqrt N$ \footnote{$\sigma$ is the standard deviation of $D^2_{min}$ and $N$ is the number of measurements.} \\
acceleration ($g$) & & & \\ \hline
$\sim$1.8 & 0.211 & 0.0275\\
1 & 0.193 & 0.0015\\
$\sim$0 & 0.137 & 0.0008 \\
\end{tabular}
\end{ruledtabular}
\end{table}

\section*{Supplemental Material: Further experiment details}

Here we give an overview of the experimental design, however, the full technical details of the hardware developed for this experiment are presented in \cite{murdoch12_IEF}. The two concentric cylinders used to make the Taylor-Couette shear cell are cast Acrylic tubes; the outer cylinder has an inner radius of 195 mm and the inner cylinder has an outer radius of 100 mm.  Both the inner and outer cylinders are 200 mm in height. The outer cylinder is fixed and its inner surface is rough with a layer of particles, the outer surface of the inner cylinder is also rough but it is free to rotate, and the floor between the two cylinders is smooth and fixed in place (see Fig. 6).  

\vspace{1cm}
\begin{figure}[h]
\centering
\includegraphics[width=0.7\columnwidth]{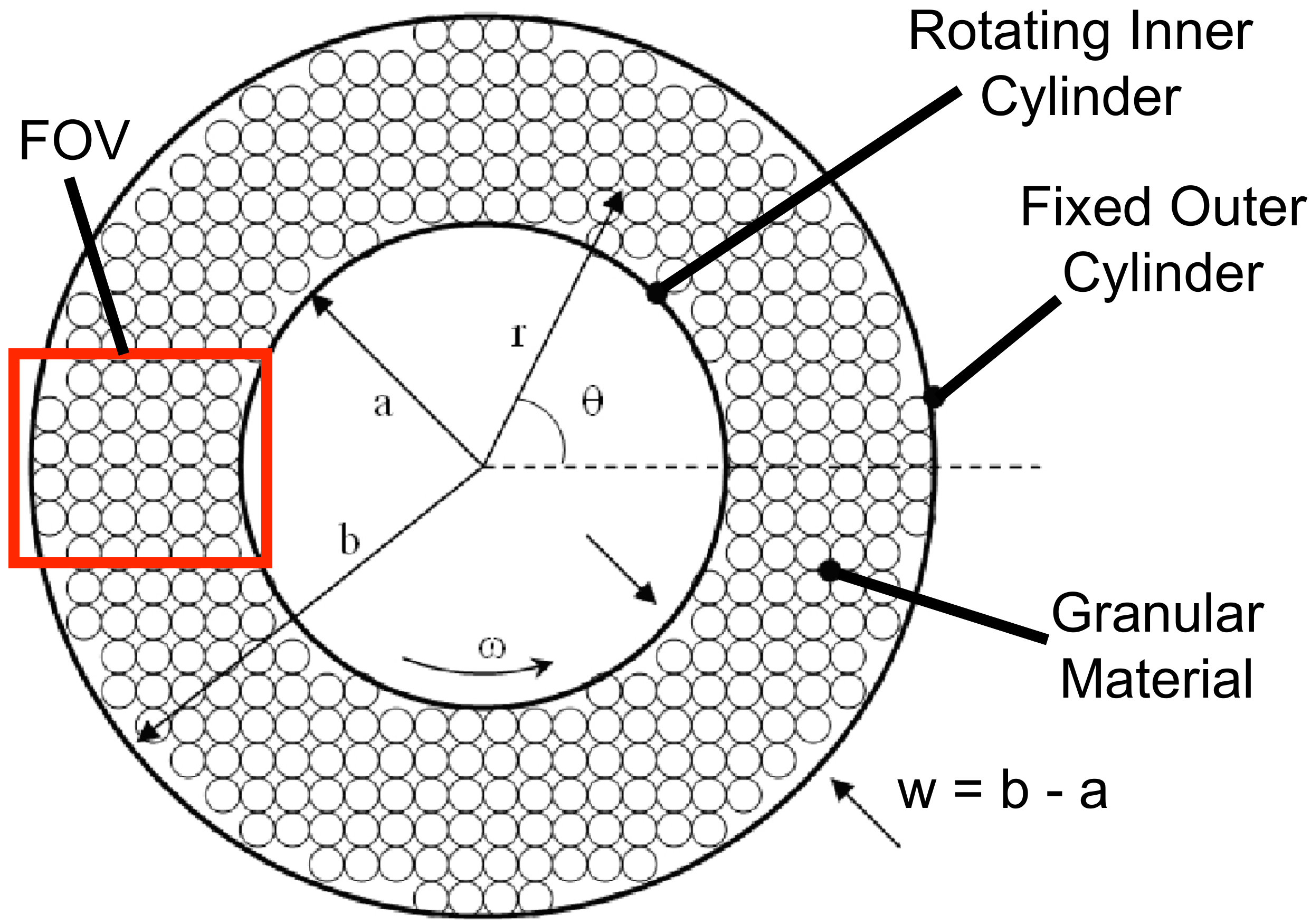}
\label{fig:TaylorCouette}
\caption{The Taylor-Couette Geometry ($a$ = inner cylinder radius, $b$ = outer cylinder radius, $w$ = width of shear region, $r$ = radial distance, $\theta$ = angular distance, $\omega$ = inner cylinder angular velocity). The camera field of view (FOV) is also shown. Image adapted from \cite{toiya06}}. 
\end{figure}

We confine the granular material by exerting a very low positive force on the top surface of the beads using a \emph{pressure plate}: a sprung loaded movable, transparent disk.  This ensures all sidewalls can sustain forces to contain the particles.  The force applied to the top surface of the beads is distributed between 3 springs; each spring provides a force of 2.2 N at the normal filling height ($\psi$ = 0.6), a force of 0 N at the minimum filling height ($\psi$ = 0.645) and a maximum force of 4.4 N at the maximum filling height ($\psi$ = 0.555; \citep{onoda90}). This is equivalent to pressure variations of between 0 and 149 Pa, assuming that the pressure is equally distributed over the entire area of the plate.

The shear cell is mounted inside the A300 Zero-G aircraft using an experiment rack. As the motor was being used only to a fraction of its maximum operating power, in combination with a large gearing ratio, the required rotation rate was reached almost instantaneously (within 0.02 seconds). The motor rotation rate was constant to within $\pm$0.01\%, and no indications of vibrations from the motor could be observed - the images of the stationary regions overlay to better than 1 pixel (193 micron) resolution. To minimise the effects of aircraft vibrations we attempt to isolate the shear cell from the aircraft by mounting silent blocks (a type of vibration isolator made of rubber) between the two strut profiles on which the shear cell is resting and the rest of the support structure frame. However, despite this attempt to isolate the shear cell, experiments during steady flight (i.e., between parabolas), during which the cameras were switched on but the motor remained off revealed that the particles are sensitive to aircraft vibrations with the largest velocity fluctations caused by the vibrations being $\sim$2 $\times$ 10$^{-2}$ mm s$^{-1}$ in the radial direction; an order of magnitude smaller than the magnitude of the measured secondary flow.

\bibliography{MurdochPRL.bib} 

\end{document}